\UseRawInputEncoding
\documentclass[11pt]{article}
\usepackage{graphicx}
\usepackage{amssymb,amsopn}
\usepackage{amsmath,mathrsfs}
\usepackage{mathtools}
\usepackage{subcaption}
\usepackage[margin=1.25in]{geometry}
\usepackage[usenames,dvipsnames]{color}
\usepackage{url}
\usepackage{flexisym}
\usepackage{comment}
\usepackage[colorlinks = true,
            linkcolor = blue,
            urlcolor  = blue,
            citecolor = blue,
            anchorcolor = blue]{hyperref}

\newcommand\pubdate{\today}

\def\Title#1{\begin{center} {\LARGE #1 } \end{center}}
\def\Author#1{\begin{center}{ \sc #1} \end{center}}
\def\Address#1{\begin{center}{ \it #1} \end{center}}

\newcommand\pubblock{
  \rightline{\begin{tabular}{l}      
         \pubdate \end{tabular}}}
\newenvironment{Abstract}{\begin{quotation} \begin{center}
                       ABSTRACT
     \end{center}\bigskip  }{\end{quotation}}

\begingroup
\catcode`\_11
\catcode`\:11
\gdef\math_bsym_Bin:Nn#1#2{%
\allowbreak\math_char:NNn 2#1{#2}\nobreak}
\gdef\math_bsym_Rel:Nn#1#2{%
\allowbreak\math_char:NNn 3#1{#2}}
\endgroup

\newcommand\disproc{\begin{center}\rule[-0.2in]{\hsize}{0.01in}\\\rule{\hsize}{0.01in}\\
    \vskip 0.1in
Presented at DIS2022: XXIX International Workshop on Deep-Inelastic Scattering and Related Subjects, Santiago de Compostela, Spain, May 2-6 2022. \\
\rule{\hsize}{0.01in}\\\rule[+0.2in]{\hsize}{0.01in} \end{center}}

\begin{document}

\pubblock

\Title{
  $W$~+~charm associated hadroproduction:\\
  relevance of Shower Monte Carlo effects
}

%\bigskip

%To switch to footnotes with special symbols:
\renewcommand{\thefootnote}{\fnsymbol{footnote}}

\Author{
  Giuseppe~Bevilacqua$^{1}$,
  Maria~Vittoria~Garzelli$^{2,}$\footnote[4]{Presenter e-mail address:
    \texttt{maria.vittoria.garzelli@desy.de}},
  Adam~Kardos$^{3}$,
  Lorant~Toth$^{3}$
}

%To switch back to footnotes with arabic numbers:
\renewcommand*{\thefootnote}{\arabic{footnote}}

%\medskip
%\smallskip

\Address{
  $^1$ ELKH-DE Particle Physics Research Group, H-4010 Debrecen, P.O. Box 105, Hungary\\
  $^2$ II. Institut f\"ur Theoretische Physik, Universit\"at
  Hamburg, Luruper Chaussee 149, D~--~22761 Hamburg, Germany\\
  $^3$ Institute of Physics, University of Debrecen, H-4010 Debrecen, P.O. Box 105, Hungary
}

\medskip

\begin{Abstract}
  \noindent
  Data on $W + D$-meson and $W + c$-jet hadroproduction have recently started to be included in at least some of the parton distribution function fits, mainly because of their potential to constrain the strange quark content of the proton. In this
  contribution
we present predictions for $W + D$-meson and $W + c$-jet production with NLO QCD accuracy matched to parton shower. We show how including the latter effects, as well as hadronization, beam remnant and multiple parton in\-te\-rac\-tion effects
present
in Shower Monte Carlo codes, is fundamental to provide consistent comparisons with the current experimental data by the ATLAS and CMS collaborations, as required for non-biased extractions of the strange and antistrange quark PDFs.
\end{Abstract}

\disproc
%\vfill
%\vfill
%\newpage
%\def\thefootnote{\fnsymbol{footnote}}
\setcounter{footnote}{0}

\paragraph{Introduction}
\label{sec:intro}

The strangeness content of the proton 
is constrained, at present, mainly thanks to
Drell-Yan (+ jets) data at the Large Hadron Collider (LHC) and older
$((\nu$+~$\bar{\nu})~+~A)$ deep inelastic scattering data, with large uncertainties~\cite{Accardi:2016ndt, Alekhin:2017olj, Faura:2020oom}. The $W$~+~single charm hadroproduction process presents important sensitivity to the strange sea, due to the fact that, among
the four
parton-level subprocesses contributing to it
 at leading order (LO),
two involve strange quarks as initial states:
$\bar{s}g~\rightarrow~W^+ \bar{c}$, $sg~\rightarrow~W^- c$.
On the other hand, the other two involve down quarks as initial states:
$\bar{d}g \rightarrow W^+ \bar{c}$, $dg \rightarrow W^- c$,
and represent an additional relevant contribution due to quark-flavour mixing, considering that
the $V_{cd}$ element of the $V_{\mathrm{CKM}}$ matrix is far from being null.
It is fundamental to include off-diagonal CKM effects, to properly estimate the $s-\bar{s}$ asymmetry. The latter is predicted by the theory~\cite{Catani:2004nc}, but its magnitude is currently unknown. Data collected so far have not pointed to a large asymmetry and many parton distribution function (PDF) fits are still obtained under the assumption of a null asymmetry. Some recent PDF fits, however, allow for an asymmetry (see e.g. Ref.~\cite{Bailey:2020ooq}).
At next-to-leading order (NLO) additional initial state channels open up, providing additional sensitivity to light quark and gluon PDFs, while limiting the direct sensitivity to $s$ and $\bar{s}$ quarks.

At hadron colliders,  $W$ + single charm hadroproduction is investigated and measured in two channels: $W~+~D$-meson and $W~+~c$-jet, with
$W$ decaying leptonically.
In both channels $W^+$ and $W^-$ events are collected and analysed separately. Experimental data were reported at both Tevatron, by the CDF~\cite{Aaltonen:2007dm, Aaltonen:2012wn} and D0~\cite{Abazov:2008qz} col\-la\-bo\-ra\-tions, and at the LHC, by the ATLAS~\cite{Aad:2014xca}, CMS~\cite{Chatrchyan:2013uja,Sirunyan:2018hde,CMS:2021oxn} and LHCb~\cite{Aaij:2015cha} collaborations.  
The fiducial cuts differ analysis by analysis. In general the $c$-jet undergoes
larger transverse momentum cuts than the $D$-meson (i.e. $p_{T,\,c-jet} > 20 - 25$~GeV vs. $p_{T,\,D} > 5 - 8$~GeV).

In the most recent analyses events are classified as either ``Opposite Sign'' (OS), when the selected $D$-meson (or $c$-jet) and charged lepton $\ell$ from $W$-boson decay have charges of opposite sign, or ``Same Sign'' (SS), when the selected $D$-meson (or $c$-jet) and
$\ell$ have charges of the same sign.
In order to minimize the backgrounds, in particular the $W^\pm c\bar{c}$ one, 
(OS - SS) fiducial inclusive and differential cross sections $d\sigma/d|\eta_\ell|$ are then provided.
 Extensive comparisons of experimental data for these ob\-ser\-va\-bles to theory predictions including NLO QCD + Shower Monte Carlo (SMC) effects, obtained through \textsc{PowHel}~\cite{Bevilacqua:2021ovq}, which uses as a basis the Powheg NLO~+~Parton Shower (PS) matching scheme~\cite{Nason:2004rx} as implemented in \textsc{Powheg box}~\cite{Alioli:2010xd}, in association with matrix elements obtained with the \textsc{Helac-NLO} package~\cite{Bevilacqua:2011xh}, and is interfaced to \textsc{Pythia8}~\cite{Sj_strand_2015} to describe SMC effects beyond the first radiative emission, are reported in Ref.~\cite{Bevilacqua:2021ovq}. The \textsc{PowHel} implementation includes quark flavour mixing effects, treats the charm quark as a massive particle, and preserves the correlations among heavy-quark masses, PDFs and $\alpha_s(M_Z)$, as well as spin-correlations in $W$-boson decays. A comparable level of data/theory agreement is achieved for both the ATLAS and CMS analyses of Ref.~\cite{Aad:2014xca,Sirunyan:2018hde}. The agreement of theory predictions with the ($W^+$ + $W^-$) data is slightly better than for the $W^-$ and $W^+$ cases considered separately.
Further distributions (e.g. $p_{T,\ell}$ and $p_{T,D}$) have started to be considered only in the most recent analyses and can be useful for testing and comparing the reliability, advantages and shortcomings of theoretical approaches using different assumptions/inputs/accuracy.

\paragraph{Relevance of SMC effects}
\label{sec:effects}

\begin{figure}
\begin{center}
  \includegraphics[width=0.51\columnwidth]{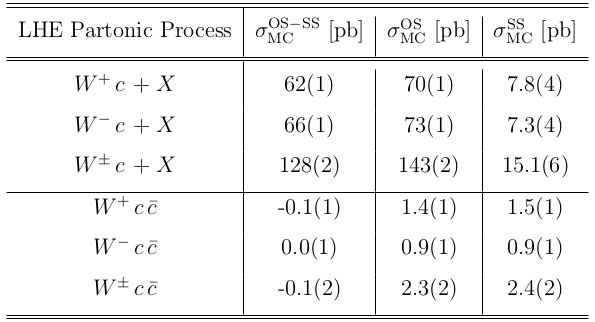}\\
  \includegraphics[width=0.44\columnwidth]{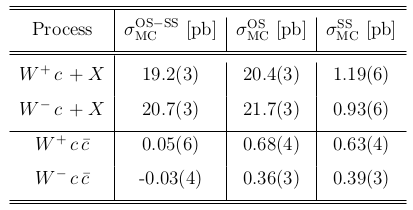}
  \includegraphics[width=0.44\columnwidth]{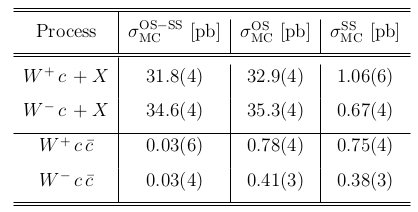}
  \caption{\label{fig:ss} Separate OS and SS contributions to the \textsc{PowHel + Pythia8}  fiducial inclusive (OS~-~SS) cross sections for the $W^\pm~+~D^*(2010)^\mp$-meson CMS analysis of Ref.~\cite{Sirunyan:2018hde} (upper panel), the $W^\pm~+~D^\mp$-meson (lower left panel) and the $W + c$-jet (lower right panel) ATLAS analysis of Ref.~\cite{Aad:2014xca}. Together with the inclusive fiducial $W^+ \bar{c} + X$ and $W^- c + X$ contributions, the contribution due to the real NLO $W^\pm c\bar{c}$ corrections already incorporated into the inclusive fiducial cross sections is also shown explicitly. See Ref.~\cite{Bevilacqua:2021ovq} for more detail.}
  \end{center}
\end{figure}

To illustrate the relevance of parton shower effects, it is worth focusing on the SS contribution to the fiducial cross section. Among the subprocesses entering $W+c$ production at NLO QCD accuracy, there are some that contribute to the SS cross section, i.e. the NLO real corrections $u\bar{d} \rightarrow W^+ c\bar{c}$ and $\bar{u}d \rightarrow W^- c\bar{c}$. Further contributions to the SS cross section come from $g~\rightarrow~c \bar{c}$ splittings in the parton shower. The latter
may amount to some percent~\cite{Bevilacqua:2021ovq}, i.e. the same order of magnitude of NNLO scale uncertainties~\cite{Czakon:2020coa}. The overall size of the SS contribution amounts to up to O(10\%) of the (OS~-~SS) cross-section, implying that one can not neglect it, i.e. one cannot approximate the SS contribution as null, when comparing theory predictions with the (OS~-~SS) experimental data. The size of the SS contribution relative to the (OS~-~SS) one depends on the analysis cuts, and is more pronounced for the $D$-meson analyses (looser $p_T$ cuts) than for the $c$-jet analyses, and for higher center-of-mass energies, as can be seen comparing among each other the tables in the panels of Fig.~\ref{fig:ss}.

\begin{figure}
  \begin{center}
  \includegraphics[width=0.70\textwidth]{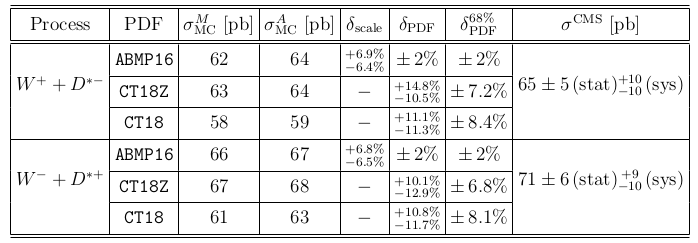}
  \includegraphics[width=0.285\textwidth]{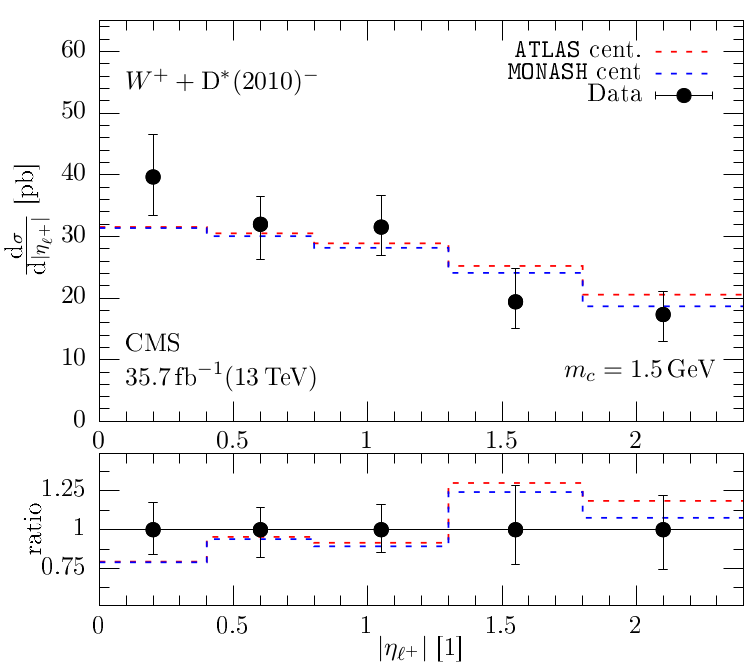}\\
  \caption{\label{fig:cms} Left: inclusive (OS~-~SS) fiducial cross sections by \textsc{PowHel + Pythia8}, using two different \textsc{Pythia8} tunes (Monash and ATLAS A14) and different PDFs as input vs. CMS $W^\mp+D^*(2010)^\pm$ experimental data.
    Right: $|\eta_{\ell^+}|$ fiducial distribution by
    \textsc{PowHel + Pythia8} with two different \textsc{Pythia8} tunes (Monash and ATLAS A14) vs. CMS $W^+~+~D^*(2010)^-$ experimental data.  See Ref.~\cite{Bevilacqua:2021ovq} for more detail.}
  \end{center}
\end{figure}

Differences between theory predictions with different tunes, corresponding to different values of the parameters encoding
multiple parton interaction (MPI) and beam remnant
effects, amount to~$\mathcal{O}$(10\%), which is the same order of magnitude of NLO scale uncertainties. This is shown in the left panel of Fig.~\ref{fig:cms}
in case of the $W^\pm+D^*(2010)^\mp$ inclusive fiducial cross sections after cuts of the CMS analysis of Ref.~\cite{Sirunyan:2018hde}  and in the right panel of Fig.~\ref{fig:cms} in case of the $d\sigma/d|\eta_{\ell^+}|$ distribution, respectively. These theory predictions and experimental data can be useful to constrain those PDFs whose uncertainty is currently larger than scale uncertainties. At NLO the uncertainty due to seven-point scale variation reaches $\sim \mathcal{O}$(10)\%, whereas at NNLO it amounts to few \%~\cite{Czakon:2020coa}.

\paragraph{Conclusions}
\label{sec:conclu}

On the one hand, properly accounting for MPI and beam remnant effects when comparing theory predictions to $W~+~c$ data in PDF fits is fundamental given the
size
of these effects, as follows from the results of our study.
On the other hand, PS effects, which reduce the (OS~-~SS) fiducial cross sections incorporating them compared to the fixed-order case, should also be accounted for~when~comparing theory predictions with experimental data in fits of PDFs at fixed order, either~by including them in theory predictions, or by correcting the data to the fixed-order level. Otherwise, when using fixed-order QCD theory predictions to compare to particle level data including parton shower effects, one risks to underestimate the strange content of the proton.
Additionally, one might reasonably expect that including effects due to NLO electroweak (EW) corrections would reduce the theoretical cross sections presented in this work by very few percent. EW effects affect also the shower development. The study of the latter is left for future work. 
For the future we foresee strange sea determinations using as a basis either theory predictions including
all the aforementioned
effects, or experimental data properly corrected for
them.

%\Acknowledgements
\paragraph{Acknowledgements}
The work of G.B., A.K. and L.T. was supported in part by COST and by the National Research, Development and Innovation Fund in Hungary, under grant K 125105.
%\textcolor{magenta}{[check if more info is available]}
The work of M.V.G. was supported in part by
%the Deutsche Forschungsgemeinschaft under contract \textcolor{magenta}{XXX} and
 the Bundesministerium f\"ur Bildung und Forschung under contracts 05H18GUCC1 and 05H21GUCCA.
%\textcolor{magenta}{[Add the DFG and BMBF contracts.]}
\\
%\\
\bibliographystyle{JHEP}
\bibliography{wcproc}  % file wcproc.bib
\end{document}